\documentclass[a4paper,12pt]{article}
\usepackage{mathrsfs}
\usepackage{graphicx}
\usepackage{amsfonts}

\newcommand{\sect}[1]{\setcounter{equation}{0}\section{#1}}

\begin{document}
\topmargin 0pt \oddsidemargin 0mm

\renewcommand{\thefootnote}{\fnsymbol{footnote}}
\begin{titlepage}
\begin{flushright}
\end{flushright}

\vspace{5mm}
\begin{center}
{\Large \bf  Holography, UV/IR Relation, Causal Entropy Bound and
Dark Energy } \vspace{12mm}

{\large
Rong-Gen Cai\footnote{Email address: cairg@itp.ac.cn}, Bin Hu\footnote{Email address: hubin@itp.ac.cn}
and Yi Zhang~\footnote{Email address: zhangyi@itp.ac.cn}\\
\vspace{8mm}
{ \em Institute of Theoretical Physics, Chinese Academy of Sciences, \\
   P.O. Box 2735, Beijing 100190, China}}
\end{center}
\vspace{5mm} \centerline{{\bf{Abstract}}}
 \vspace{5mm}
The constraint on the total energy in a given spatial region is
given from holography by the mass of a black hole which just fits in
that region, which leads to an UV/IR relation: the maximal energy
density in that region is proportional to $M_p^2/L^2$, where $M_p$
is the Planck mass and $L$ is the spatial scale of that region under
consideration. Assuming the maximal black hole in the universe is
formed through gravitational collapse of perturbations in the
universe, then the ``Jeans" scale of the perturbations gives a
causal connection scale $R_{\rm CC}$. For gravitational
perturbations, $R^{-2}_{\rm CC}={\rm Max}(\dot H+2H^2, -\dot H)$ for
a flat universe. We study the cosmological dynamics of the
corresponding vacuum energy density by choosing the causal
connection scale as the IR cutoff in the UV/IR relation,  in the
cases of the vacuum energy density as an independently conserved
energy component and an effective dynamical cosmological constant,
respectively. It turns out that only the case with the choice
$R_{\rm CC}^{-2}= \dot H +2H^2$, could be consistent with the
current cosmological observations when the vacuum density appears as
an independently conserved energy component. In this case, the model
is called holographic Ricci scalar dark energy model in the
literature.

\end{titlepage}

\newpage
\renewcommand{\thefootnote}{\arabic{footnote}}
\setcounter{footnote}{0} \setcounter{page}{2}
\sect{Introduction} Since the discovery of accelerating expansion of
the universe by observing distant supernova~\cite{Riess,Perl}, the
nature of dark energy has been one of the hottest issues in
cosmology and theoretical physics. Now a lot of astronomical
observations indicate that the energy budget of the universe
consists of approximately $4\%$ baryon matter, $23\%$ dark matter,
$73\%$ dark energy and negligible radiation. The simplest and
economic way to explain the accelerating expansion of the universe
is due to a tiny positive cosmological constant introduced first by
Einstein himself in 1917. It is well-known that the cosmological
constant and the vacuum expectation value of some quantum fields
 are undistinguished. Thus, the cosmological constant acting
as the solution of the dark energy suffers from the so-called fine
tuning problem: what is the physical mechanism that sets the value
of the cosmological constant to its current observed value, which is
120 orders of magnitude smaller than the naive theoretical
expectation. Also there exists the so-called coincidence problem for
dark energy: why does the cosmological constant dominate the
universe just recently. In other words, the coincidence problem can
also be expressed as follows. Why the energy densities of dark
energy and dark matter are comparable just recently?

Since the cosmological constant is entangled with the vacuum
expectation value of some quantum fields, the cosmological constant
problem therefore is essentially an issue of quantum gravity.
Indeed, general relativity together with quantum field theory could
shed some lights on the cosmological constant problem.

It is widely believed and precisely tested that particle physics
can be accurately described by an effective field theory with an
ultraviolet (UV) cutoff less than the Planck mass $M_p$, provided
that all momenta and field strengths are small compared with this
cutoff to the appropriate power~\cite{Cohen}. This is indeed the
case in the absence of gravity.  When gravity effect is taken into
account, however, something very strange appears. Black hole
thermodynamics tells us that a  black hole has an entropy
proportional to its horizon area. In Einstein gravity theory,
black hole entropy satisfies the so-called area formula, $S=A/4G$,
where $A$ is the horizon area of the black hole and $G$ is the
Newtonian constant. One learns from statistical physics that
entropy of a system describes the number of microscopic degrees of
freedom of the system, and that entropy is an extensive quantity
and is always proportional to volume of the system. For a black
hole, entropy is proportional to its area. This implies that for a
gravity system, its effective degrees of freedom are drastically
reduced, compared to the same system without gravity. This
indicates that the underlying theory describing the nature must be
not a local quantum field theory. In other words, there exists a
range of validity for a local effective field theory to describe a
system with gravity. This leads to it unbelievable that the
cosmological constant should be in the order of Planck mass,
obtained from the naive estimate of local effective field theory
without taking into account the gravity effect.

To realize this argument, Cohen {\it et al.}~\cite{Cohen} proposed
a relation between UV and infrared (IR) cutoffs. To be
self-contained, let us briefly repeat some key steps to the
relation. For an effective quantum field theory confined in a box
of size $L$ with UV cutoff $\Lambda$, the entropy should scale
extensively as $S \sim L^3\Lambda^3$. However, black hole
thermodynamics leads Bekenstein to argue that the maximal entropy
of the box with volume $L^3$ scales as its area, instead of the
volume $L^3$.  The Bekenstein entropy bound could be satisfied for
an effective local quantum field theory if the following
inequality is obeyed
\begin{equation}
\label{1eq1} L^3\Lambda^3 < S_{\rm BH} = \pi L^2 M_p^2,
\end{equation}
where $S_{\rm BH}$ is the entropy of a black hole with horizon
radius $L$. One can see from (\ref{1eq1}) that the length $L$
acting as an IR cutoff is no longer independent of the UV cutoff,
but scales as $M_p^2/\Lambda^3$.  However, as argued by Cohen {\it
et al.},  there is evidence that conventional quantum field fails
at an entropy well below the bound (\ref{1eq1}). They gave a more
stronger constraint on the IR cutoff, which excludes all states
residing within their Schwarzschild radius. Note that the maximal
energy density in the effective theory is $\Lambda^4$, the mass in
a box with volume $L^3$ is $\Lambda^4 L^3$. Assuming that the mass
is less than the mass of a black hole with radius $L$ leads to the
following constraint
\begin{equation}
\label{1eq2}
 \Lambda^4 L^3 \leq L M_p^2,
 \end{equation}
 where the IR cutoff scales as $M_p/\Lambda^2$. This bound is more
 restrictive than (\ref{1eq1}). To see this, let us consider the case
 where (\ref{1eq2}) is nearly saturated. In that case, the entropy
 is $S_{\rm max} \simeq S_{\rm BH}^{3/4}$, which is less than the
 black hole entropy $S_{\rm BH}$.

 The UV/IR relation (\ref{1eq2}) leads to a very interesting consequence on the
 cosmological constant problem. If the effective local quantum field
 theory is valid in an arbitrarily large volume up to the Planck
 mass, the contribution of the vacuum energy density to the
 cosmological constant is $\sim (10^{19} {\rm Gev})^4$. If SUSY
 exists and is broken at energy scale $\sim {\rm Tev}$, then the contribution of the vacuum energy density to the
 cosmological constant is $\sim ( {\rm Tev})^4$. On the other hand,
 if the bound (\ref{1eq2}) plays some role,  then the contribution of the vacuum energy density to the
 cosmological constant is $M_p^2/L^2$. If choosing the IR cutoff
 as the current horizon size of the universe, one has
 $\Lambda^4 \sim M_p^2/L^2 \sim (10^{-3} {\rm ev})^4$. This is
 exactly in the same order as the observed dark energy scale.

 However, as found by Hsu~\cite{Hsu}, if one takes the current Hubble
 horizon as the IR cutoff in (\ref{1eq2}), although the energy
 density $\rho \sim \Lambda^4$ can match the dark energy density of
 the universe, it cannot make the universe accelerating expansion since its equation of
 state is the same as the one for dark matter. Li~\cite{Li} found
 that the particle horizon of the universe also cannot take the
 job, instead the even horizon of the universe acting as the IR cutoff
 can derive the universe to accelerating expand, and the vacuum energy density in
 this case can fit the data well. However, even
 horizon is a global concept of spacetime, the event horizon of the
 universe is determined by future evolution of the universe. As a
 result, it is not easy to understand why the current dark energy
 density is determined by future evolution of the universe, rather
 than the past of the universe~\cite{Cai}. In the paper~\cite{Cai},
 combining general relativity and uncertainty relation in quantum
 mechanics, we argued that the energy density of quantum
 fluctuations of spacetime could act as the dark energy currently
 observed. The dark energy density is characterized by the age of
 the universe, and could be consistent with astronomical data if the
 unique numerical parameter in this model is taken to be a number of
 order unity. But it was found that it is not consistent with the
 evolution history of our universe that there is a matter dominated
 decelerated phase in the past. Several ways out have been proposed
 such as considering interaction between dark energy and dark
 matter~\cite{WeiCai1}, and replacing the age of the universe by the
 conformal time of the universe~\cite{WeiCai2}, etc.
 For further considerations see \cite{others}, for example.

 In this paper, considering black hole in the universe is
 formed by gravitational collapse of perturbations of cosmological
 spacetime and the ``Jeans" length of the perturbations sets a
 causal connection scale, beyond which black hole cannot formed
 very likely, we study the cosmological dynamics of the vacuum
 energy density by use of the causal connection scale as the IR
 cutoff in (\ref{1eq2}). Here we would like to mention that in fact,
 the UV/IR relation (\ref{1eq2}) does not resolve the cosmological
 constant problem, since as a pure constant energy, black hole cannot
 form without fluctuations of energy.

\sect{Holography and Causal Entropy Bound}

Given a closed system with fixed energy $E$, which fits in a sphere
with radius $R$ in three spatial dimensions, what is the maximal
entropy of the system? Bekenstein was the first to consider this
issue. Based on black hole thermodynamics, he argued there exists an
upper bound on the entropy of the system~\cite{Bek}
\begin{equation}
\label{2eq1}
 S \le S_{\rm B}=2\pi ER=\pi M_P^2 R_gR,
 \end{equation}
 where $R_g=2E/M_p^2$ is the Schwarzschild radius of the system.
  This bound is called Bekenstein entropy bound. This bound is
 believed to be universal valid for a system with limited
 self-gravity, which means that the gravitational self-energy is
 negligibly small compared to its total energy $E$. However, it is
 interesting to note that the bound is saturated even for a four
 dimensional Schwarzschild black hole which is a strongly
 self-gravitating object (note that it is no longer saturated
 for higher dimensional ($D>4)$ Schwarzschild black holes).
 In addition, it is worth mentioning here
 that although the Bekenstein bound (\ref{2eq1}) is derived from
 black hole thermodynamics and generalized second law, it is
 independent of gravitational theory and spacetime
 dimensions~\cite{CMO}.

 When taking into account the effect of gravity, based on the black
 hole entropy relation with horizon area, the so-called
 entropy-area relation in Einstein gravity, it is argued that the
 maximal entropy of a system is bounded by its area $A$~\cite{Holo}
 \begin{equation}
 \label{2eq2}
 S \le S_{\rm H}= M_p^2 A/4.
 \end{equation}
 That is, the maximal entropy of a system is given by entropy of the black
 hole  with the same size as the system.  The entropy bound
 (\ref{2eq2}) is called holographic entropy bound. For a limited
 self-gravitating system, its Schwarzschild radius $R_g <R$, the
 holographic entropy bound (\ref{2eq2}) is less restrictive than
 the Bekenstein bound (\ref{2eq1}). For both entropy bounds, they
 are all given by the size of a space-like surface enclosing the
 system under consideration. Then it is interesting to see whether the entropy bounds
 (\ref{2eq1}) and (\ref{2eq2}) can be applicable to our universe and what
 consequences can be acquired from those entropy bounds.
 Bekenstein himself generalized the entropy bound (\ref{2eq1}) to
 the cosmological setting by replacing $R$ by the particle horizon
 in a FRW universe. On the other hand, Fischler and
 Susskind~\cite{FS}
 proposed that the area of the particle horizon should give an
 bound of matter entropy on the backward-looking
light cone in the form (\ref{2eq2}). However, it is easy to see
that this version of entropy bound could be violated in a closed
universe. Several proposals have been suggested in order to remedy
this problem, for example, to replace the particle horizon by
Hubble horizon or apparent horizon~\cite{gene}. Generalizing the
concept of the light-sheet proposed by Fischler and Susskind,
Bousso~\cite{Bousso} suggested the covariant entropy bound, which
is applicable to arbitrary spacetimes. The covariant entropy bound
gives an entropy bound on a light-like hypersurface. Therefore, in
order to give an entropy bound on a space-like region, a
``space-like projection" has to be performed.

It is interesting to note that there exists an improved covariant
entropy bound, which is applicable to entropy on space-like
hypersurfaces and pasts several critical tests, proposed by Brustein
and Veneziano~\cite{BV}(for a recent review see \cite{Brus}). The
improved covariant entropy bound is called causal entropy bound. For
a system with limited self-gravitating, the Bekenstein bound is the
tightest, while in other situations, the causal entropy bound is
argued to be a strongest one. The causal entropy bound is given as
follows. Consider a generic space-like hypersurface, defined by the
equation $\tau=0$, and a compact region lying within it defined by
$\sigma \le 0$, the entropy contained in this region, $S(\tau=0,
\sigma \le 0)$, is bounded by $S_{\rm CEB}$
\begin{eqnarray}
\label{2eq3}
 S_{\rm CEB} &=& l_p^{-2}\int_{\sigma <0}d^4x \sqrt{-g}
 \delta(\tau)  \sqrt{{\rm Max}_{\pm} [(G_{\mu\nu}\pm
 R_{\mu\nu})\partial^{\mu}\tau \partial^{\nu}\tau]}  \nonumber \\
 &=& l_p^{-1}\int_{\sigma <0}d^4x \sqrt{-g}\delta(\tau)
  \sqrt{[{\rm Max}_{\pm}[(T_{\mu\nu}\pm T_{\mu\nu}\mp g_{\mu\nu}T/2)
  \partial^{\mu}\tau \partial^{\nu}\tau]},
  \end{eqnarray}
  where $l_p$ is the Planck length, $G_{\mu\nu}$ and $R_{\mu\nu}$
  are Einstein tensor and Ricci tensor, respectively, $T_{\mu\nu}$
 is stress energy tensor of matter and $T$ its trace.  In the
 second equality, the Einstein equations $G_{\mu\nu}=8\pi G
 T_{\mu\nu}$ have been used. The causal entropy bound (\ref{2eq3})
 is manifestly covariant and invariant under reparametrization of
 the hypersurface equation, while the reality of $S_{\rm CEB}$ is
 assured if the source matter obeys the weak energy condition,
 $T_{\mu\nu}\partial^{\mu}\tau \partial^{\nu}\tau \ge 0$, since
 the sum of two combinations in (\ref{2eq3}) and thus their
 maximum, are positive.

 Here we are not interested in the causal entropy bound itself
 (\ref{2eq3}), but the motivation which leads to the causal entropy
 bound. Note that both the entropy bound (\ref{1eq1}) and
 holographic entropy bound (\ref{2eq2}) are given by assuming the
 entropy in a given region
 of space is bounded by entropy of a largest black hole which can
 fit in that region, while the bound (\ref{1eq2}) is given by the
 mass of the largest black hole fitting in that region. The Hubble
 entropy bound in cosmology~\cite{gene} is given by assuming the
 largest black hole in the universe is the one with horizon radius of
 Hubble horizon. However, note that gravitational collapse happens
 within only ``Jeans" length of gravitational fluctuations in a
 universe, and perturbations with wavelength beyond the ``Jeans"
 length are causally disconnected. The causal entropy bound is
 just based on the argument that black hole with larger radius
 than the ``Jeans" length cannot formed very likely in the cosmological
 setting~\cite{BV}. Then the remained problem is to find out the
 causal connection (CC) scale $R_{\rm CC}$.

 The authors of \cite{BV} identified the causal connection scale
 $R_{\rm CC}$ for a FRW universe as follows. In the Hamiltonian
 approach~\cite{hamil}, the Fourier components of a (normalized) perturbation
 in a FRW universe and of its (normalized) conjugate momentum
 satisfy the Schroedinger-like equations
 \begin{eqnarray}
 \label{2eq4}
 \hat  \Psi_k '' +[k^2 -(z^{1/2})''z^{-1/2}]\hat \Psi_k=0, \nonumber\\
 \hat \Pi_k '' +[k^2-(z^{-1/2})''z^{1/2}]\hat \Pi_k=0,
 \end{eqnarray}
 where $k$ is the comoving momentum, a prime stands for derivative
 with respect to conformal time, and $z^{1/2}$ is the so-called
 ``pump field", a combination of the various backgrounds which
 depends on the special perturbation under study. These
 perturbation equations clearly indicate a ``Jeans-like" CC
 comoving momentum
 \begin{eqnarray}
 \label{2eq5}
 k^2_{\rm CC} &=& {\rm Max}[(z^{1/2})''z^{-1/2}, (z^{-1/2})''z^{1/2}]
 \nonumber \\
  &=& {\rm Max} [{\cal K}'+{\cal K}^2, -{\cal K}'+{\cal K}^2],
 \end{eqnarray}
 where ${\cal K}=(z^{1/2})'z^{-1/2}$. Since the tensor
 perturbation is always present in any case, it is therefore natural to
 consider the tensor perturbation as the ``pump field" $z^{1/2}$.
 In that case, $z^{1/2}$ is given by the scale factor $a$, so that
 one has ${\cal K}=a'/a$. Note that the comoving momentum $k$
 gives a definition of a proper ``Jeans" CC length $R_{\rm
 CC}=ak^{-1}_{\rm CC}$, and the latter can be further expressed as
\begin{equation}
\label{2eq6} R^{-2}_{\rm CC}={\rm Max}[\dot H +2H^2, -\dot H],
\end{equation}
where the dot stands for derivative with respect to the cosmic time
and $H$ is the Hubble parameter of the universe. The result
(\ref{2eq6}) is valid for a flat FRW universe. It turns out that for
a FRW universe with any spatial curvature, the CC scale (\ref{2eq6})
is changed to~\cite{BV}
\begin{equation}
\label{2eq7} R^{-2}_{\rm CC}={\rm Max}[\dot H +2H^2+\kappa/a^2,
-\dot H+ \kappa/a^2],
\end{equation}
where $\kappa$ is the spatial curvature of the universe. Brustein
and Veneziano arrived at the causal entropy bound (\ref{2eq3})
starting from the CC length scale (\ref{2eq7}). The essence of the
causal entropy bound is that the largest black hole in the
universe is the one with horizon radius given by $R_{\rm CC}$ in
(\ref{2eq7}).

Assuming the matter source in the FRW universe is a perfect fluid
with energy-momentum tensor $T^{\mu}_{\nu}={\rm diag}(\rho_t, p_t,
p_t, p_t)$, and with the help of the $00$ components of the Ricci
tensor and Einstein tensor, $R_{00}=-3 (\dot H +H^2)$ and $G_{00}=3
(H^2 +\kappa/a^2)$, the CC scale (\ref{2eq7}) can be further written
as~\cite{BV}
\begin{eqnarray}
\label{2eq8} R^{-2}_{\rm CC} &=&\frac{1}{3}{\rm Max}_{\mp}(G_{00}\mp
R_{00}) \nonumber \\
&=& 4\pi M_p^{-2} \ {\rm Max}( \rho_t/3-p_t, \rho_t +p_t) \nonumber \\
&=& 4\pi M_p^{-2} \rho_t\ {\rm Max}[ (1/3-\omega_t), (1+\omega_t)]
\end{eqnarray}
In the third equality, we have used the equation of state of the
perfect fluid, $p_t=\omega_t \rho_t$. It is clear from (\ref{2eq8})
that the first term is larger if $\omega_t<-1/3$, while the second
term larger as $\omega_t>-1/3$. For the current universe,
astronomical observations indicate that the first term is larger
than the second term. This implies that the first term is more
suitable for as the IR cutoff for the universe at present. This will
be shown indeed the case shortly.

Now we would like to see the consequence by choosing the CC scale
(\ref{2eq7}) as the IR cutoff in the UV/IR relation (\ref{1eq2}).
For simplicity, we will consider the case of a flat universe in what
follows and parameterize the vacuum energy density (\ref{1eq2}) as
\begin{equation}
\label{2eq9} \rho_{\Lambda}=\frac{3c^2 m_p^2}{R_{\rm CC}^{2}},
\end{equation}
by introducing a parameter $c^2$, where $m_p$ is the reduced Planck
mass. Obviously, if $\dot H \ll H^2$, or $|\dot H| \sim H^2$, the
vacuum energy density (\ref{2eq9}) gives the current observed dark
energy density if the parameter $c^2$ is of order unity.

\sect{Dynamics of holographic vacuum energy}

In this section, for completeness, we will separately discuss the
cosmological evolution of the holographic vacuum energy by
choosing different CC scales in (\ref{2eq6}).  Also let us  first
notice that the vacuum energy density (\ref{2eq9}) could appear in
the Friedmann equation in two different forms: (1) The vacuum
energy density obeys the continuity equation acting as an
independent component of energy budget of the universe. That is,
it obeys
\begin{equation}
\label{3eq1}
 \dot \rho_{\Lambda} +3 H (1+\omega_{\Lambda})\rho_{\Lambda}=0,
 \end{equation}
 where $\omega_{\Lambda}$ is the equation of state of the vacuum
 energy density. In that case, there is no interaction between the
 vacuum energy and other sources like dark matter in the universe.
 In this case the vacuum energy density will be called independent
 vacuum energy model. (2) The vacuum energy density (\ref{2eq9})
 could also appear as a dynamical cosmological
 constant. That is, its equation of state is always
 $\omega_{\Lambda}=-1$. In this case, due to the Bianchi identity,
 there must exist some interaction between the vacuum energy and
 dark (and baryon) matter $\rho_m$ (in this paper we will neglect the
 contribution of radiation in the universe). The total energy
 obeys the continuity equation
 \begin{equation}
 \label{3eq2}
 \dot \rho_m +\dot \rho_{\Lambda} +3 H \rho_m=0,
 \end{equation}
 where we have used the assumption $\rho_{\Lambda}+p_{\Lambda}=0$.
 In the following, we will consider separately the two cases.

\subsection{IR Cutoff 1: $R_{\rm CC}^{-2}=\dot H +2H^2$}

Let us first notice that the Ricci scalar of a flat FRW universe is
$R=6 (\dot H +2H^2)$. In this case, the vacuum energy density
(\ref{2eq9}) is proportional to the Ricci scalar curvature. Such a
holographic dark energy model is introduced first by Gao {\it et
al.}~\cite{Gao} without mentioning their motivation. Here we stress
that the Ricci scalar curvature gives a causal connection scale of
perturbations in the universe. In order to be self-contained, here
we give some key results.

(1) {\it Independent vacuum energy model}.
 The Friedmann equation
reads
\begin{equation}
\label{3eq3}
 H^2 = \frac{1}{3m_p^2}(\rho_m +\rho_{\Lambda}).
\end{equation}
Substituting $\rho_{\Lambda}=3c^2m_p^2 (\dot H + 2H^2) $ into
(\ref{3eq3}), one can obtain
\begin{equation}
\label{3eq4}
 h^2= \Omega_{m0}e^{-3x} +
 \frac{c^2\Omega_{m0}}{2-c^2}e^{-3x}+c_0e^{-(4-2/c^2)x},
 \end{equation}
 where $h=H/H_0$, $x=\ln a$, $c_0$ is an integration constant and
 $\Omega_{m0}$ is the current fraction dark matter energy density.
  Clearly the integration constant $c_0$ has to satisfy the
constraint
\begin{equation}
\label{3eq5}
 \Omega_{m0} +\frac{c^2\Omega_{m0}}{2-c^2} +c_0 =1.
 \end{equation}
Further, the second and third terms in (\ref{3eq4}) can be viewed
as
 the fraction vacuum density
 \begin{equation}
 \tilde
 \rho_{\Lambda}=\frac{c^2\Omega_{m0}}{2-c^2}e^{-3x}+c_0e^{-(4-2/c^2)x}.
 \end{equation}
 Using (\ref{3eq1}), one can get the equation of state for the
 vacuum energy density as
 \begin{eqnarray}
 \label{3eq7}
 \omega_{\Lambda} &=& -1 -\frac{\tilde \rho_{\Lambda}'}{3 \tilde
 \rho_{\Lambda}}, \nonumber \\
    &=& -1 + \frac{\frac{c^2}{2-c^2}\Omega_{m0}+\frac{1}{3}\left(4-\frac{2}{c^2}\right)c_0 e^{(2/c^2-1)x}}
    {\frac{c^2}{2-c^2}\Omega_{m0}+c_0e^{(2/c^2-1)x}},
    \end{eqnarray}
 where a prime stands for derivative with respect to $x$. The
 current equation of state is given by
 \begin{equation}
 \label{3eq8}
\omega_{\Lambda0}= -1 +
\frac{\frac{c^2}{2-c^2}\Omega_{m0}+\frac{1}{3}\left(4-\frac{2}{c^2}\right)c_0}
    {\frac{c^2}{2-c^2}\Omega_{m0}+c_0}.
    \end{equation}
In Fig.~1 we plot the equation of state for the vacuum energy
density. It clearly shows that in early time it behaves as a dust
matter, while it behaves like a phantom field at late time. In
addition, we can see from (\ref{3eq7}) that at infinite future, if
$c^2<2$,
\begin{equation}
\label{3eq9}
 \omega_{\Lambda\infty}= \frac{1}{3} -\frac{2}{3c^2}.
 \end{equation}
$\omega_{\Lambda\infty} <-1$ provided $c^2<1/2$. It could be a
reliable dark energy model. For further discussions on this model,
see, for example, \cite{Ricci}. In addition, let us mention here
that the authors of \cite{Nojiri} considered the case of a
combination of Hubble parameter, event horizon, particle horizon and
the life time of the universe (if finite) as an IR cutoff; Medved in
a footnote of \cite{Medved}  mentioned the possibility of the causal
connection scale as the scale of ``causal boundary".

\begin{figure}[htb]
\centering
\begin{minipage}[c]{.58\textwidth}
\centering
\includegraphics[width=\textwidth]{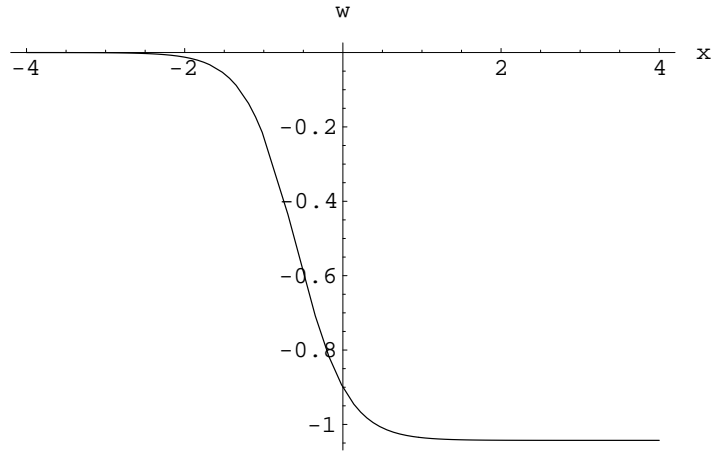}
\caption{This plot shows the equation of state for the vacuum
energy density versus $x=\ln a$, provided $\omega_{\Lambda0}=-0.9$
and $ \Omega_{m0}=0.3$. In this case, $c_0=0.604$ and $c^2=0.484$}
\end{minipage}
\end{figure}

(2) {\it Dynamical cosmological constant}. In this case, our
starting point is the two equations (\ref{3eq2}) and (\ref{3eq3}).
One has
\begin{equation}
\label{3eq10} (1-2c^2) H^2= (c^2-2/3)\dot H.
\end{equation}
We then have the solution
 \begin{equation}
 \label{3eq11}
 a= a_0  (t_0 +
 \alpha t)^{1/\alpha},
 \end{equation}
 where $a_0$ and $t_0$ are two integration constants, and $\alpha=
3(2c^2-1)/(3c^2-2)$. The equation of state for the total energy is
\begin{equation}
\omega_t = -1+\frac{2}{3}\alpha.
\end{equation}
Note that in this case, due to $\rho_m \sim -\dot H$, in order to
keep the positivity of the matter energy density, one has to have
$\dot H <0$, which implies that $\omega_t >-1$ or $\alpha >0$.

 i) When $c^2>2/3$, one has $\alpha>0$, but $1/\alpha <1$.  In this
 case, the universe always decelerated expands.

 ii) When $1/2 <c^2 <2/3$, one has $\alpha <0$ and $\omega_t <-1$.
     This is not a physical allowed case.

 iii) When $1/3 <c^2 <1/2$, one has $\alpha >0$ and $1/\alpha >1$,
  the universe always accelerating expands with a power-law form.
  This implies that there is no decelerated phase in this case.
  This is not consistent with current observational fact.

 iv) When $c^2 <1/3$, one has $\alpha >0$ and $1/\alpha <1$. In this
 case, the universe always in a decelerated phase.

 As a result, if acting as an effective dynamical cosmological
 constant, the vacuum energy density is not consistent with the
 current observation data.

\subsection{IR Cutoff 2: $R_{\rm CC}^{-2}=-\dot H $}

In this case, note that in order  $ R_{\rm CC}^{-2}$ to be positive,
$\dot H $ should be negative. Let us first discuss the case as an
independent energy component.

(1) {\it Independent vacuum energy model}. In that case, the
corresponding Friedmann equation can be rewritten as
\begin{equation}
\label{3eq13}
 h^2 = \Omega_{m0} e^{-3x} -\frac{c^2}{2} (h^2)'.
 \end{equation}
 Integrating this equation yields
 \begin{equation}
 \label{3eq14}
 h^2=\Omega_{m0} e^{-3x} +c_0 e^{-2x/c^2}
 -\frac{3c^2}{3c^2-2}\Omega_{m0}e^{-3x},
 \end{equation}
 where $c_0$ is an integration constant, which should obey the
 constraint
 \begin{equation}
 \Omega_{m0} +c_0 -\frac{3c^2}{3c^2-2}\Omega_{m0}=1.
 \end{equation}
 On the other hand, the fraction vacuum energy density
 \begin{equation}
 \tilde \rho_{\Lambda}=c_0 e^{-2x/c^2}
 -\frac{3c^2 }{3c^2-2}\Omega_{m0}e^{-3x},
 \end{equation}
 can give its equation of state
 \begin{equation}
 \omega_{\Lambda}= -1 +\frac{\frac{2c_0}{3c^2}e^{-2x/c^2}
 -\frac{3c^2}{3c^2-2}\Omega_{m0}e^{-3x}}{c_0
 e^{-2x/c^2}-\frac{3c^2}{3c^2-2}\Omega_{m0}e^{-3x}}.
 \end{equation}
 The current equation of state is
 \begin{equation}
 \omega_{\Lambda0}=-1 +\frac{\frac{2c_0}{3c^2}
 -\frac{3c^2}{3c^2-2}\Omega_{m0}}{c_0
 -\frac{3c^2}{3c^2-2}\Omega_{m0}},
 \end{equation}
 and at infinite future $x \to \infty$,
 \begin{equation}
 \omega_{\Lambda \infty}=-1 +\frac{2}{3c^2},
\end{equation}
provided $c^2 > 2/3$.  In Fig.~2  we plot the equation of state for
the vacuum energy density provided $\Omega_{m0}=0.3$ and
$\omega_{\Lambda0}=-0.9$. The equation of state diverges at some
time in the past. Fig.~3 plots the fraction Hubble parameter squared
$h^2$, which turns to be negative at some time in the past. Clearly
this is not a physical solution. To see that this case could not be
consistent with evolution history of the universe, let us look at
the second and third terms in (\ref{3eq14}). In order to have an
accelerating expansion, one has $c^2>1$ from the second term, while
one has to have $c^2>2/3$ if requiring the second term is dominant
over the third term currently. Then one has $3c^2/(3c^2-2)>1$ and
the second term would be dominant, which always leads to a negative
$h^2$ in the early time.

\begin{figure}[htb]
\centering
\begin{minipage}[c]{.58\textwidth}
\centering
\includegraphics[width=\textwidth]{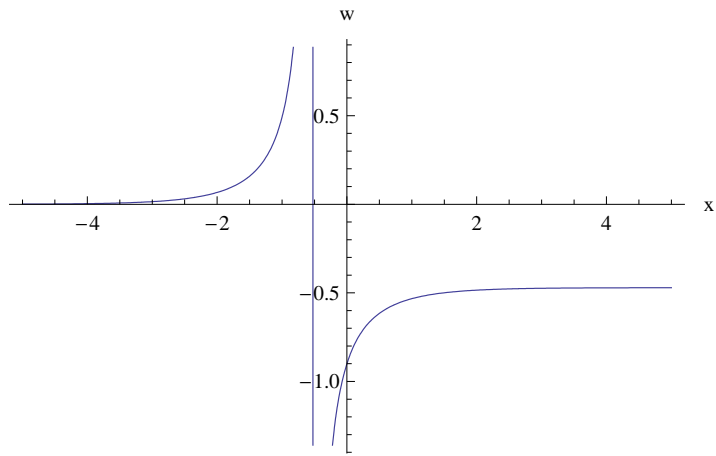}
\caption{This plot shows the equation of state for the vacuum energy
density versus $x=\ln a$, provided $\omega_{\Lambda0}=-0.9$ and $
\Omega_{m0}=0.3$.}
\end{minipage}
\end{figure}

\begin{figure}[htb]
\centering
\begin{minipage}[c]{.58\textwidth}
\centering
\includegraphics[width=\textwidth]{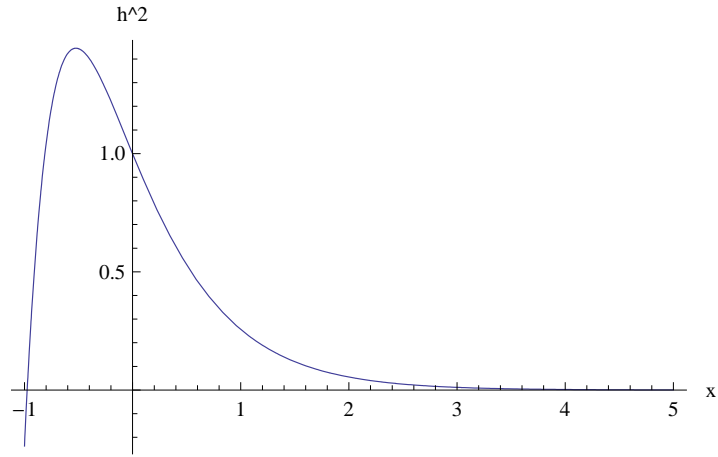}
\caption{This plot shows the fraction Hubble parameter squared
$h^2$ versus $x=\ln a$, provided $\omega_{\Lambda0}=-0.9$ and $
\Omega_{m0}=0.3$.}
\end{minipage}
\end{figure}

(2) {\it Dynamical cosmological constant}. In this case, the
Friedmann equation can be cast to
\begin{equation}
H^2 =-(c^2 +2/3)\dot H,
\end{equation}
which has the solution of the scale factor
\begin{equation}
a=a_0 (t_0 +\beta t)^{1/\beta},
\end{equation}
where $1/\beta= c^2+2/3$, while the total equation of state is
\begin{equation}
\omega_t= -1 +\frac{2}{3}\beta.
\end{equation}
Clearly, in this case, the universe always accelerating
(decelerated ) expands as $c^2 >1/3$ ($c^2 < 1/3$). This is again
not consistent with current observational data.

\sect{Conclusions}

 From holographical property of gravity, one has the so-called UV/IR
 relation. The dark energy problem is an IR problem, while the cosmological constant
 problem is an UV problem.  It is
 therefore natural to make a connection between the UV/IR relation
 and the dark energy problem. The casual entropy bound for a spatial region in a
 cosmological setting is given by assuming the maximal black hole
 in the universe is formed by gravitational collapse with the ``Jeans" scale
 of perturbations, beyond which black hole cannot form very likely.
 Therefore the ``Jeans" scale of perturbations in the universe
 naturally leads to an IR cutoff in the cosmological setup.

The causal connection scale is given by $R_{\rm CC}^{-1}=\sqrt{{\rm
Max}(\dot H +H^2,-\dot H)}$ for gravitational perturbation in a FRW
universe~\cite{BV}. We studied the cosmological dynamics of the
vacuum energy density by choosing the causal connection scale as the
IR cutoff in the UV/IR relation, in the cases of $R_{\rm
CC}^{-2}=\dot H+2H^2$ and $R_{\rm CC}^{-2}=-\dot H$, respectively.
Also we separately considered the cases of the corresponding vacuum
density as an independently conserved energy component and as an
effective dynamical cosmological constant. It turns out only the
case with the choice $R^{-2}_{\rm CC}= \dot H+2H^2$ could be
consistent with current cosmological data if it acts as the observed
dark energy. This model is called holographic Ricci scalar model in
the literature since $R^{-2}_{\rm CC}$ is proportional to the Ricci
scalar of the FRW spacetime in this case. As a result, it appears
that the causal connection scale acts as a new IR cutoff in the
cosmological setting. It is of some interesting to investigate other
cosmological consequences for this model. Finally let us stress that
our discussions do not exclude some interaction between the vacuum
energy density and dark matter if $\Omega_{\Lambda} \ne -1$ in
(\ref{3eq2}).

\section*{Acknowledgments}
This work was supported in part by a grant from the Chinese Academy
of Sciences with No. KJCX3-SYW-N2,  grants from NSFC with No.
10821504 and No. 10525060.


\begin{thebibliography}{99}
\bibitem{Riess}A.~G.~Riess {\it et al.}  [Supernova Search Team Collaboration],
  Astron.\ J.\  {\bf 116}, 1009 (1998)
  [arXiv:astro-ph/9805201].
\bibitem{Perl} S.~Perlmutter {\it et al.}  [Supernova Cosmology Project Collaboration],
  Astrophys.\ J.\  {\bf 517}, 565 (1999)
  [arXiv:astro-ph/9812133].

\bibitem{Cohen}A.~G.~Cohen, D.~B.~Kaplan and A.~E.~Nelson,
  Phys.\ Rev.\ Lett.\  {\bf 82}, 4971 (1999)
  [arXiv:hep-th/9803132].

\bibitem{Hsu}S.~D.~H.~Hsu,
  Phys.\ Lett.\  B {\bf 594}, 13 (2004)
  [arXiv:hep-th/0403052].

\bibitem{Li}M.~Li,
  Phys.\ Lett.\  B {\bf 603}, 1 (2004)
  [arXiv:hep-th/0403127].
\bibitem{Cai} R.~G.~Cai,
  Phys.\ Lett.\  B {\bf 657}, 228 (2007)
  [arXiv:0707.4049 [hep-th]].

\bibitem{WeiCai1}  H.~Wei and R.~G.~Cai,
  arXiv:0707.4052 [hep-th];
 H.~Wei and R.~G.~Cai,
  Phys.\ Lett.\  B {\bf 655}, 1 (2007)
  [arXiv:0707.4526 [gr-qc]];
 I.~P.~Neupane,
  arXiv:0708.2910 [hep-th].


\bibitem{WeiCai2} H.~Wei and R.~G.~Cai,
  Phys.\ Lett.\  B {\bf 660}, 113 (2008)
  [arXiv:0708.0884 [astro-ph]];
H.~Wei and R.~G.~Cai,
  Phys.\ Lett.\  B {\bf 663}, 1 (2008)
  [arXiv:0708.1894 [astro-ph]].


\bibitem{others}X.~Wu, Y.~Zhang, H.~Li, R.~G.~Cai and Z.~H.~Zhu,
  arXiv:0708.0349 [astro-ph];
Y.~Zhang, H.~Li, X.~Wu, H.~Wei and R.~G.~Cai,
  arXiv:0708.1214 [astro-ph];
M.~Maziashvili,
  Phys.\ Lett.\  B {\bf 666}, 364 (2008)
  [arXiv:0708.1472 [hep-th]];
  K.~Y.~Kim, H.~W.~Lee and Y.~S.~Myung,
  Phys.\ Lett.\  B {\bf 660}, 118 (2008)
  [arXiv:0709.2743 [gr-qc]];
I.~P.~Neupane,
  Phys.\ Rev.\  D {\bf 76}, 123006 (2007)
  [arXiv:0709.3096 [hep-th]];
M.~Maziashvili,
  Phys.\ Lett.\  B {\bf 663}, 7 (2008)
  [arXiv:0712.3756 [hep-ph]];
 J.~Zhang, X.~Zhang and H.~Liu,
  Eur.\ Phys.\ J.\  C {\bf 54}, 303 (2008)
  [arXiv:0801.2809 [astro-ph]].
Y.~W.~Kim, H.~W.~Lee, Y.~S.~Myung and M.~I.~Park,
  arXiv:0803.0574 [gr-qc].
J.~P.~Wu, D.~Z.~Ma and Y.~Ling,
  Phys.\ Lett.\  B {\bf 663}, 152 (2008)
  [arXiv:0805.0546 [hep-th]].


\bibitem{Bek}J.~D.~Bekenstein,
  Phys.\ Rev.\  D {\bf 23}, 287 (1981).


 \bibitem{CMO}  R.~G.~Cai, Y.~S.~Myung and N.~Ohta,
  Class.\ Quant.\ Grav.\  {\bf 18}, 5429 (2001)
  [arXiv:hep-th/0105070];
R.~G.~Cai and Y.~S.~Myung,
  Phys.\ Lett.\  B {\bf 559}, 60 (2003)
  [arXiv:hep-th/0210300].


 \bibitem{Holo}G.~'t Hooft,
  arXiv:gr-qc/9310026;
L.~Susskind,
  J.\ Math.\ Phys.\  {\bf 36}, 6377 (1995)
  [arXiv:hep-th/9409089].

\bibitem{FS}W.~Fischler and L.~Susskind,
  arXiv:hep-th/9806039.

  \bibitem{gene}R.~Easther and D.~A.~Lowe,
  Phys.\ Rev.\ Lett.\  {\bf 82}, 4967 (1999)
  [arXiv:hep-th/9902088];
  G.~Veneziano,
  Phys.\ Lett.\  B {\bf 454}, 22 (1999)
  [arXiv:hep-th/9902126];
  G.~Veneziano,
  arXiv:hep-th/9907012;
D.~Bak and S.~J.~Rey,
  Class.\ Quant.\ Grav.\  {\bf 17}, L83 (2000)
  [arXiv:hep-th/9902173];
N.~Kaloper and A.~D.~Linde,
  Phys.\ Rev.\  D {\bf 60}, 103509 (1999)
  [arXiv:hep-th/9904120].



\bibitem{Bousso}R.~Bousso,
  JHEP {\bf 9907}, 004 (1999)
  [arXiv:hep-th/9905177];
  R.~Bousso,
  JHEP {\bf 9906}, 028 (1999)
  [arXiv:hep-th/9906022];
R.~Bousso,
  Rev.\ Mod.\ Phys.\  {\bf 74}, 825 (2002)
  [arXiv:hep-th/0203101].


\bibitem{BV}R.~Brustein and G.~Veneziano,
  Phys.\ Rev.\ Lett.\  {\bf 84}, 5695 (2000)
  [arXiv:hep-th/9912055].

\bibitem{Brus}R.~Brustein,
  Lect.\ Notes Phys.\  {\bf 737}, 619 (2008)
  [arXiv:hep-th/0702108].

\bibitem{hamil}R.~Brustein, M.~Gasperini and G.~Veneziano,
  Phys.\ Lett.\  B {\bf 431}, 277 (1998)
  [arXiv:hep-th/9803018];
  A.~Ghosh, G.~Pollifrone and G.~Veneziano,
  Phys.\ Lett.\  B {\bf 440}, 20 (1998)
  [arXiv:hep-th/9806233].


\bibitem{Gao}C.~Gao, X.~Chen and Y.~G.~Shen,
  arXiv:0712.1394 [astro-ph].

\bibitem{Ricci}C.~J.~Feng,
  arXiv:0806.0673 [hep-th];
  C.~J.~Feng,
  Phys.\ Lett.\  B {\bf 670}, 231 (2008)
  [arXiv:0809.2502 [hep-th]];
C.~J.~Feng,
  arXiv:0810.2594 [hep-th];
L.~N.~Granda and A.~Oliveros,
  Phys.\ Lett.\  B {\bf 669}, 275 (2008)
  [arXiv:0810.3149 [gr-qc]].
L.~N.~Granda and A.~Oliveros,
  arXiv:0810.3663 [gr-qc];
 E.~N.~Saridakis,
  Phys.\ Lett.\  B {\bf 660}, 138 (2008)
  [arXiv:0712.2228 [hep-th]];
E.~N.~Saridakis,
  JCAP {\bf 0804}, 020 (2008)
  [arXiv:0712.2672 [astro-ph]];
E.~N.~Saridakis,
  Phys.\ Lett.\  B {\bf 661}, 335 (2008)
  [arXiv:0712.3806 [gr-qc]];
 L.~Xu, W.~Li and J.~Lu,
  arXiv:0810.4730 [astro-ph];
  C.~J.~Feng,
  arXiv:0812.2067 [hep-th];
K.~Y.~Kim, H.~W.~Lee and Y.~S.~Myung,
  arXiv:0812.4098 [gr-qc].
\bibitem{Nojiri}S.~Nojiri and S.~D.~Odintsov,
  Gen.\ Rel.\ Grav.\  {\bf 38}, 1285 (2006)
  [arXiv:hep-th/0506212].
\bibitem{Medved}A.~J.~M.~Medved,
  arXiv:0802.1753 [hep-th].

\end{thebibliography}
\end{document}